\documentclass{master-rv9x6}

\usepackage{makeidx}

\makeindex

\usepackage{latexsym,amsfonts}

\usepackage{graphicx,epic,amssymb,longtable}

\def\a{\alpha}
\def\b{\beta}

\def\d{\delta}
\def\e{\eta}

\def\l{\lambda}

\def\D{\Delta}
\def\m{\mu}
\def\n{\nu}
\def\r{\rho}

\def\s{\sigma}

\def\e{\varepsilon}

\def\pa{\partial}
\def\rg{{\rm g}}
\def\be{\begin{equation}}
\def\ee{\end{equation}}
\def\beq{\begin{eqnarray}}
\def\eeq{\end{eqnarray}}
\def\nn{\nonumber}
\def\ca{{\cal A}}

\def\ch{{\cal H}}
\def\cl{{\cal L}}
\def\cm{{\cal M}}
\def\cn{{\cal N}}

\def\cv{{\cal V}}
\def\cV{{\cal V}}

\def\RR{{\mathbb{R}}}

\def\ZZ{{\mathbb{Z}}}
\def\E{{E_{10}}}
\def\KE{{K(E_{10})}}

\def\Bt{{\tilde B}}

\newcommand{\ft}[2]{{\textstyle {\frac{#1}{#2}} }}
\newcommand{\tr}{{\rm tr \,}}

\newcommand{\frakg}{{\mathfrak{g}}}
\newcommand{\frakh}{{\mathfrak{h}}}
\newcommand{\frakn}{{\mathfrak{n}}}
\newcommand{\frakk}{{\mathfrak{k}}}

\newcommand{\Sl}{{\mathfrak{sl}}}
\newcommand{\SD}{\sum_{\a\in\Delta_+}}
\newcommand{\Smult}{\sum_{s=1}^{{\rm mult}(\a)}}
\newcommand{\mult}{{{\rm mult}\,}}
\newcommand{\Ref}[1]{(\ref{#1})}
\newcommand{\non}{\nonumber\\}

\newcommand{\bqn}{\begin{eqnarray}}\newcommand{\eqn}{\end{eqnarray}}
\renewcommand{\theequation}{\thesection.\arabic{equation}}

\begin{document}

\chapter{GRAVITATIONAL BILLIARDS, DUALITIES AND\\
HIDDEN SYMMETRIES}

\markboth{H. Nicolai}{Gravitational Billiards, Dualities and 
Hidden Symmetries}

\vspace*{-18pt}   

\author{H. Nicolai}
\addcontentsline{toc}{author}{H. Nicolai}

\address{Max-Planck-Institut f\"ur Gravitationsphysik,
Albert-Einstein-Institut,\\ 
Am M\"uhlenberg 1, D-14476 Golm, Germany}

\begin{abstract}
The purpose of this article is to highlight the fascinating, but only very 
incompletely understood relation between Einstein's theory and its 
generalizations on the one hand, and the theory of indefinite, and in 
particular hyperbolic, Kac Moody algebras on the other. The elucidation 
of this link could lead to yet another revolution in our understanding of 
Einstein's theory and attempts to quantize it.
\end{abstract}

\section{Introduction}
\setcounter{equation}{0}
\setcounter{theorem}{0}
\setcounter{lemma}{0}

As we look back 90 years to take stock of what has been achieved since
Einstein explained gravity in terms of spacetime geometry and its 
curvature, the progress is impressive. Even more impressive is the
wealth of structure contained in Einstein's equations which has been 
revealed by these developments. Major progress has been made concerning  

\begin{itemize}
\item Exact solutions (Schwarzschild, Reissner-Nordstr\"om, Kerr,
      axisymmetric stationary solutions,...)
\item Cosmological applications (standard FRW model of cosmology,
      inflationary universe,...)
\item Mathematical developments (singularity theorems, black hole
      uniqueness theorems, studies of the initial value problem,
      global stability results,...)
\item Conceptual developments (global structure and properties of 
      space-times, horizons, black hole entropy, quantum
      theory in the context of cosmology, holography,...)
\item Canonical formulations (Dirac's theory of constrained systems,
      ADM formalism, Ashtekar's variables,...)
\item Higher dimensions (Kaluza Klein theories, brane worlds,...)
\item Unified theories `beyond' Einstein (supergravity, superstrings, 
      supermembranes and M(atrix) theory,...)
\item Quantizing gravity (perturbative and canonical
      quantization, path integral approaches, dynamical triangulations,
      spin networks and spin foams,...) 
\end{itemize}

All these subjects continue to flourish and are full of promise for 
further and exciting developments (hinted at by the dots in the 
above list). No doubt many of them will be discussed and elaborated 
in other contributions to this volume. In this article, we will 
concentrate on yet another line of research that evolved out of the 
study of Einstein's equations and its locally supersymmetric extensions, 
and which points to another deep, and still mostly unexplored property 
of Einstein's theory. It may well be that the main discoveries in this 
direction still remain to be made, and that, once they have been made, 
they will also have a profound impact on attempts to quantize Einstein's 
theory (or some bigger theory containing it). This is the subject of

\begin{itemize}
\item Hidden symmetries and dualities
\end{itemize}

The first hint of these symmetries appeared in Ref.~\refcite{Buchdahl}, 
where a transformation between two static solutions of Einstein's
equations was given, which in modern parlance is nothing but a
$T$-duality transformation. A decisive step was Ehlers' discovery in 1957 
of a solution generating symmetry,\cite{Ehlers} nowadays known as the 
`Ehlers $SL(2,\RR)$ symmetry' which acts on certain classes of solutions
admitting one Killing vector. In 1970, R. Geroch demonstrated the 
existence of an {\em infinite dimensional} extension of the Ehlers 
group acting on solutions of Einstein's equations with two commuting 
Killing vectors (axisymmetric stationary solutions).\cite{Geroch} In the 
years that followed, the Geroch group was extensively studied by general 
relativists with the aim of developing `solution generating techniques'
(see\cite{Hoenselars,Kramer} and references therein for an {\em entr\'ee}
into the literature). The field received new impetus with the discovery 
of `hidden symmetries' in supergravities, most notably the exceptional 
$E_{7(7)}$ symmetry of maximal $N=8$ supergravity.\cite{CJ} These results 
showed that the Ehlers and Geroch groups were but special examples of
a more general phenomenon.\cite{Julia,Julia1,BM,BGM,Nic1} With the shift
of emphasis from solution generating techniques to the Lie algebra and 
the group theoretical structures underlying these symmetries, it became
clear that the Geroch symmetry is a non-linear and non-local realization 
of an affine Lie group (a loop group with a central extension), with 
Lie algebra $A_1^{(1)} = \widehat{\Sl (2,\RR)}_{ce}$. This completed
earlier results by general relativists who had previously realized 
`half' of this affine symmetry in terms of `dual potentials'.\cite{KC} 
Likewise, generalizations of Einstein's theory, and in particular its 
locally supersymmetric extensions were shown to possess similar infinite 
dimensional symmetries upon reduction to two dimensions. These results 
also provided a direct link to the integrability  of these theories 
in the reduction to two dimensions, i.e. the existence of Lax pairs 
for the corresponding equations of motion.\cite{Maison,BZ,BM}

All these duality invariances of Einstein's theory and its extensions
apply only to certain truncations, but do not correspond to properties 
of the {\em full} theory, or some extension thereof. Our main point 
here will be to review and highlight the evidence for even larger 
symmetries which would {\em not} require any truncations, and whose 
associated Lie algebras belong to a special class of infinite dimensional
Lie algebras, namely the so-called {\em indefinite Kac Moody 
Algebras}.\cite{Kac,MP,GO} We will discuss two examples of such Lie 
algebras, namely the rank three algebra $AE_3$,\cite{FF} which is
related to Einstein's theory in four dimensions, and secondly (but only
very briefly), the maximal rank 10 algebra $\E$, which is singled out from
several points of view, and which is related to maximal $D=11$
supergravity.\cite{CJS} We can thus phrase the central open 
question as follows:

\begin{center}
\begin{flushleft}
{\it Is it possible to extend the known duality symmetries of Einstein's
equations to the full non-linear theory without any symmetry reductions?}
\end{flushleft}
\end{center}

\noindent A perhaps more provocative, way to pose the question is

\begin{center}
{\it Is Einstein's theory integrable?} 
\end{center}

In this form, the question may indeed sound preposterous to anyone 
with even only a passing familiarity with the complexities
of Einstein's equations, which are not only the most beautiful, 
but also the most complicated partial differential equations 
in all of known mathematical physics. What is meant here, however, is 
not the usual notion of integrability in the sense that one should be 
able to write down the most general solution in closed form. Rather, 
it is the `mappability' of the initial value problem for Einstein's theory,
or some M theoretic extension thereof, onto a group theoretical structure 
that itself is equally intricate, and so infinite in extent that we may 
never be able to work it out completely, although we know that it exists. 
Even a partial answer to the above question would constitute a 
great advance, and possibly clarify other unsolved problems of general
relativity. To name but one: the `conserved charges' associated
with these Lie algebras would almost certainly be linked to the
so far elusive `observables' of pure gravity (which might better be 
called `perennials'\cite{Kuchar}) -- which we believe should exist,
though no one has ever been able to write down a single one explicitly. 

Last but not least, duality symmetries have come to play a prominent role 
in modern string theory in the guise of $T,S$ and $U$ dualities, where 
they may provide a bridge to the non-perturbative sector of the theory 
(see\cite{GPR,HT,OP} and references therein). Here, we will not dwell 
too much on this side of the story, however, because the duality groups 
considered in string theory so far are descendants of the {\em finite
dimensional} Lie groups occurring in $D\geq 4$ supergravity, whereas
here we will be mostly concerned with the {\em infinite dimensional}
symmetries that emerge upon reduction to $D\leq 2$ dimensions, and
whose role and significance in the context of string theory are not
understood. Still, it seems clear that infinite dimensional symmetries
may play a key role in answering the question, what M Theory -- the
conjectural and elusive non-perturbative and background independent
formulation of superstring theory -- really is, because that question
may well be closely related (or even equivalent) to another one,
namely

\begin{center}
{\it What is the symmetry underlying M Theory?} 
\end{center}
\noindent

There has been much discussion lately about the maximally extended hyperbolic 
Kac Moody algebra $\E$ as a candidate symmetry underlying M Theory, {\it i.e.} 
$D=11$ supergravity and the other maximally supersymmetric theories related 
to IIA and IIB superstring theory, see 
\cite{DH3,DaHeNi02,KleNic04,DN,KleNicB,DN05}, and \cite{BGH,BGGH}. 
A conceptually different proposal was made in \cite{West}, and further 
elaborated in \cite{West1,West11,West2}, according to which it is the 
`very extended' indefinite KM algebra $E_{11}$ that should be viewed as the 
fundamental symmetry ($E_{11}$ contains $\E$, but is no longer hyperbolic, 
but see \cite{GOW} for a discussion of such `very extended algebras'). 
A `hybrid' approach for uncovering the symmetries of $M$-theory combining 
\cite{DaHeNi02} and \cite{West} has been adopted in \cite{EHTW,Englert03}. 
Although our focus here is mostly on pure gravity in four space-time 
dimensions and its associated algebra $AE_3$, we will very briefly  
mention these developments in the last section.

Whatever the outcome of these ideas and developments will be, the very 
existence of a previously unsuspected link between two of the most 
beautiful concepts and theories of modern physics and mathematics,
respectively --- Einstein's theory of gravity on the one hand, and the 
theory of indefinite and hyperbolic Kac Moody algebras on the other --- 
is most remarkable and surely has some deep significance.

\section{Known duality symmetries}
We first review the two types of duality symmetries of Einstein's 
theory that have been known for a long time. The first concerns the 
{\em linearized} version of Einstein's equations and works in any 
space-time dimension. The second is an example of a {\em non-linear}
duality, which works only for the special class of solutions admitting 
two commuting Killing vectors (axisymmetric stationary and colliding 
plane wave solutions). This second duality is more subtle, not only in 
that it is non-linear, but in that it is linked to the appearance of 
an {\em infinite dimensional symmetry}.

\subsection{Linearized duality}
The duality invariance of the linearized Einstein equations generalizes
the well known duality invariance of electromagnetism in four spacetime 
dimensions. Recall that Maxwell's equations {\em in vacuo}
\be
\pa^\m F_{\m\n} = 0  \qquad , \qquad \pa_{[\m} F_{\n\rho]} = 0
\ee
are invariant under $U(1)$ rotations of the complex field strength
\be
{\cal F}_{\mu\nu} := F_{\mu\nu} + i \tilde{F}_{\m\n}
\ee
with the dual (`magnetic') field strength
\be
\tilde{F}_{\m\n} := \frac12 \epsilon_{\m\n\rho\sigma} F^{\rho\sigma}
\ee
The action of this symmetry can be extended to the combined electromagnetic 
charge $q= e + ig$, where $e$ is the electric, and $g$ is the magnetic charge.
The partner of the one-form electric potential $A_\mu$ is a {\em dual 
magnetic} one-form potential $\tilde{A}_\mu$, obeying
\be
\tilde{F}_{\m\n} := \pa_\m \tilde{A}_\n - \pa_\n \tilde{A}_\m
\ee
Observe that this dual potential can only be defined {\em on-shell},
when $F_{\m\n}$ obeys its equation of motion, which is equivalent
to the Bianchi identity for $\tilde{F}_{\m\n}$. Consequently, the $U(1)$ 
duality transformations constitute an {\em on-shell symmetry} because 
they are valid only at the level of the equations of motion. The two 
potentials $A_\mu$ and $\tilde{A}_\mu$ are obviously {\em non-local} 
functions of one another. Under their exchange, the equations of motion 
and the Bianchi identities are interchanged. Moreover, the equations of 
motion and the Bianchi identities can be combined into a single equation 
\be
\pa^\mu {\cal F}_{\mu\nu} = 0
\ee

Analogous duality transformations to the electromagnetic ones exhibited 
above exist for $p$-form gauge theories in arbitrary spacetime dimensions 
$D$ (these theories are always {\em abelian} for $p>1$). More precisely, 
an `electric'  $p$-form potential $A_{\mu_1\dots\mu_p}$ is dual to a 
`magnetic' $(D-p-2)$ potential $\tilde{A}_{\mu_1\dots\mu_{D-p-2}}$. 
A prominent example is the 3-form potential of $D=11$ 
supergravity,\cite{CJS} with a dual 6-form magnetic potential.
Upon quantization, the duality becomes a symmetry relating the weak
and strong coupling regimes by virtue of the Dirac quantization
condition $eg= 2\pi i\hbar$. This is one of the reasons why such 
dualities have recently acquired such an importance in string theory,
and why they are thought to provide an inroad into the non-perturbative
structure of the theory.

Does there exist a similar symmetry for Einstein's equations?
Remarkably, for {\em linearized} Einstein's equations in arbitrary 
space-time dimension $D$ the answer is {\em yes}.
\cite{Curtright,OPR,Hull,West,DaHeNi02,Bekeart,DT} 
However, this answer will already illustrate the difficulties one 
encounters when one tries to extend this symmetry to the full theory. 
To exhibit it, let us expand the metric as
$g_{\m\n} = \eta_{\m\n} + h_{\m\n}$, where $\eta_{\m\n}$ is the Minkowski 
metric\footnote{It is noteworthy that the construction given below appears 
to work only for a flat Minkowskian background.}, and the linearized 
fluctuations $h_{\m\n}$ are assumed to be small so we can neglect 
higher order terms. The linearized Riemann tensor is
\be
R^L_{\m\n\r\s} (h) = \pa_\m \pa_\r h_{\n\s} - \pa_\n \pa_\r h_{\m\s} - 
          \pa_\m \pa_\s h_{\n\r} + \pa_\n \pa_\s h_{\m\r} 
\ee
The linearized Einstein equations therefore read
\be
R^L_{\m\n} (h) = \pa^\r \pa_\r h_{\m\n} - \pa_\m \pa^\r h_{\r\n}
             - \pa_\n \pa^\r h_{\r\m} + \pa_\m \pa_\n {h^\r}_\r = 0
\ee
where indices are raised and lowered by means of the flat background 
metric $\eta^{\mu\nu}$. To reformulate thes equations in analogy with
the Maxwell equations in such a way that $R^L_{\m\n}=0$ gets 
interchanged with a Bianchi identity, we define  
\be
C_{\m\n | \r} := \pa_\m h_{\n\r} - \pa_\n h_{\m\r}
\ee
This `field strength' is of first order in the derivatives like the Maxwell 
field strength above, but it is {\em not} invariant under the linearized 
coordinate transformations
\be
\d h_{\mu\nu} = \pa_\mu \xi_\nu + \pa_\nu \xi_\mu \quad \Longrightarrow \quad
\delta C_{\m\n|\r} = \pa_\r (\pa_\mu\xi_\n - \pa_\n \xi_\m) \neq 0 
\ee
This is a first difficulty: unlike ordinary gauge theories, Einstein's 
theory needs {\em two} derivatives for gauge covariance.

The `Bianchi identity' now reads
\be\label{C1}
\pa_{[\m} C_{\n\r]|\s} = 0 \quad ;
\ee
and is obviously different from the usual Bianchi identity on the
Riemann tensor $R^L_{[\mu \nu\, \rho]\s} =0$. The linearized Einstein 
equations are now recovered from the equation of motion
\be\label{C2}
\pa^\m C_{\m\n|\r} = 0
\ee
if we impose the gauge condition
\be
{C_{\m\n}}^\n = 0
\ee
(imposing this condition is possible precisely because $C_{\m\n|\r}$ is 
{\em not} gauge invariant). Noticing that \Ref{C1} and \Ref{C2} are 
completely analogous to Maxwell's equations, we now introduce the `dual 
field strength'
\be
\tilde{C}_{\m_1\dots \m_{D-2}|\n} = {\e_{\m_1 \dots \m_{D-2}}}^{\r\s}
                                  C_{\r\s|\n}
\ee
It is then easy to see that vanishing divergence for one of the field 
strengths implies vanishing curl for the other, and vice versa. Furthermore,
\be\label{Ct}
\tilde{C}_{[\m_1\dots \m_{D-2}|\n]} = 0 \quad \Longleftrightarrow\quad
{C_{\m\n}}^\n = 0
\ee

On shell, where $\pa_{[\mu_1} \tilde{C}_{\mu_2\dots \mu_{D-2}]|\nu} =0$,
we can therefore introduce a `dual graviton field' 
$\tilde{h}_{\m_1\dots \m_{D-3}|\n}$, analogous to the dual `magnetic' 
potential $\tilde{A}_\mu$, with associated `field strength'
\be\label{dualG}
\tilde{C}_{\m_1\dots \m_{D-2}|\n} :=
\pa_{[\m_1} \tilde{h}_{\m_2\dots \m_{D-2}]|\n} 
\ee
Let us stress that this dual `field strength' exists only {\em on-shell}, 
i.e. when the linearized Einstein equations are satisfied. 
The tracelessness condition \Ref{Ct} requires
\be
\tilde{h}_{[\m_1\dots \m_{D-3}|\n]} = 0
\ee
This is a second new feature {\it vis-\`a-vis} Maxwell and $p$-form 
gauge theories: for $D\geq 5$, the dual graviton field transforms 
in a {\em mixed Young tableau representation}. The associated
gauge transformations are also more involved, as the gauge
parameters may likewise belong to non-trivial representations.~\cite{Bekeart} 

It does not appear possible to extend this duality invariance to the full 
non-linear theory in any obvious way. A generalization does not appear 
to exist even at first non-trivial order beyond the linear approximation, 
at least not in a way that would be compatible order by order with the 
background Lorentz invariance of the free theory. More succinctly, the No-Go 
Theorem of \cite{Bekeart} asserts that there exists no continuous deformation 
of the free theory with these properties. On the other hand, experience
has taught us that there is no No-Go Theorem without a loophole! So
we simply interpret this result as evidence that one must search in 
a different direction, giving up one or more of the seemingly
`natural' assumptions that went into its proof. An example
how one might possibly evade these assumptions is the one-dimensional 
`geodesic' $\sigma$-model over infinite dimensional cosets which will
be introduced in section~6, and which renounces manifest space-time 
Lorentz invariance.

\subsection{A nonlinear duality: the Geroch group}
Unlike for the free spin-2 theory discussed in the foregoing section,
there does exist a version of Einstein's theory possessing a non-linear 
and non-local duality symmetry, but it suffers from a different
limitation: it works only when Einstein's theory is dimensionally
reduced to two space or space-time dimensions, i.e. in the presence of
two commuting Killing vectors. For definiteness, we will take the
two Killing vectors to be spacelike, and choose coordinates such that
they are (locally) given by $\pa/\pa y$ and $\pa / \pa z$: this means
that the symmetry acts on solutions depending on two of the four
spacetime coordinates, namely $(t,x)$. In a suitable gauge we can then
write the line element as\cite{Kramer} 
\be\label{metric} ds^2 =
\D^{-1}\l^2 ( -dt^2 + dx^2) + (\r^2 \D^{-1} + \D \Bt^2) dy^2 + 2\D \Bt
dy\, dz + \D dz^2 \; , \ee 
where the metric coefficients depend only on $(t,x)$. The metric
coefficient $\Bt$ is the third component of the Kaluza Klein vector
field $(B_\mu , B_2) \equiv (0,0,\Bt)$ that would arise in the
reduction of Einstein's theory to three dimensions. The metric ansatz
\Ref{metric} can be further simplified by switching to Weyl canonical
coordinates where $\r$ is identified with the time coordinate 
\be 
\r
(t,x) = t 
\ee 
This particular choice is adapted to cosmological solutions, where $t\geq
0$ with the singularity `occurring' on the spacelike hypersurface at
time $t=0$. This is also the physical context in which we will
consider the gravitational billiards in the following
section.\footnote{If the Weyl coordinate $\rho$ is taken to be
spacelike, we would be dealing with a generalization of the so-called
Einstein-Rosen waves.}

Here we will not write out the complete Einstein equations for 
the metric ansatz \Ref{metric} (see, however,\cite{Kramer,BM,Nic1}) 
but simply note that upon dimensional reduction, the fields $(\D,\Bt)$ 
with $\D\geq 0$ coordinatize a homogeneous $\s$-model manifold 
$SL(2,\RR)/SO(2)$.\cite{NK} The equation for $\Bt$ reads 
\be
\pa_\mu ( t^{-1}\D^{2} \pa^\mu \Bt) = 0
\ee
with the convention, in this subsection only, that $\m,\n = 0,1$.
Because in two dimensions, every divergence-free vector field 
can be (locally) rewritten as a curl, we can introduce the dual 
`Ehlers potential' $B(t,x)$ by means of
\be\label{dualB0}
t\D^{-2}\pa_\mu B = \epsilon_{\mu\nu} \pa^\nu \Bt
\ee
The Ehlers potential obeys the equation of motion
\be
\pa_\mu ( t\D^{-2} \pa^\mu B) = 0
\ee
The combined equations of motion for $\D$ and $B$ can be 
compactly assembled into the so-called Ernst equation\cite{Kramer}
\be
\D \pa_\mu (t\pa^\mu {\cal E}) = t \pa_\mu {\cal E} \pa^\mu {\cal E}
\ee
for the complex Ernst potential ${\cal E} := \D + iB$. The pair $(\D,B)$
again parametrizes a coset space $SL(2,\RR)/SO(2)$, but different from 
the previous one.

To write out the non-linear action of the two $SL(2,\RR)$ symmetries,
one of which is the Ehlers symmetry, we use a notation that is
already adapted to the Kac Moody theory in the following chapters.
The relation to the more familiar `physicist's
notation' for the $SL(2,\RR)$ generators is given below: 
\be 
e \sim J^+ \;\; , \quad f \sim J^- \;\; , \quad h\sim J^3 
\ee 
In writing the variations of the fields, we will omit the infinitesimal 
parameter that accompanies each transformation. The Ehlers group is 
generated by \cite{BM,NN}
\beq\label{Ehlers} 
e_3 (\D) = 0 \quad &,&
\quad e_3(B) = -1 \non h_3(\D) = -2\D \quad &,& \quad h_3 (B) = - 2B
\non f_3 (\D) = 2\D B \quad &,& \quad f_3(B) = B^2 - \D^2 
\eeq 
The other $SL(2,\RR)$, often referred to as the Matzner Misner group, is
generated by 
\beq\label{MatznerMisner} 
e_2 (\D) = 0 \quad &,& \quad
e_2(\Bt) = -1 \non h_2(\D) = 2\D \quad &,& \quad h_2 (\Bt) = - 2\Bt
\non f_2 (\D) = -2\D \Bt \quad &,& \quad f_2(\Bt) = \Bt^2 -
\frac{\rho^2}{\D^2} 
\eeq 
(the numbering of the generators has been introduced in accordance with 
the numbering that will be used later in section~5). The Geroch group is 
now obtained by intertwining the two $SL(2,\RR)$ groups, that is by 
letting the Ehlers group act on $\Bt$, and the Matzner Misner group 
on $B$, and by iterating this procedure on the resulting `dual potentials'. 
It is not difficult to see that, in this process, one `never comes back' 
to the original fields, and an infinite tower of dual potentials is 
generated.\cite{Geroch} The Geroch group is then realized on this 
infinite tower; when projecting down this action onto the original 
fields, one ends up with a non-linear and non-local realization 
of this group.

The mathematical proof that the Lie algebra underlying the Geroch 
group is indeed $A_1^{(1)} \equiv \widehat{\Sl (2,\RR)}_{ce}$ proceeds 
by verification of the bilinear relations (no summation on $j$)\cite{NN,Nic}
\be\label{Serre1} 
[e_i , f_j]
= \d_{ij} h_j \; , \quad  [h_i, e_j] =  A_{ij} e_j \; , \quad
[h_i, f_j] = - A_{ij} f_j 
\ee
for $i,j = 2,3$, with the (Cartan) matrix 
\be\label{Geroch}
\hspace{1.5cm} A_{ij} = \left( \begin{matrix} 2 &-2\cr
                              -2 & 2  \end{matrix} \right)
\ee
The subscript `$ce$' on $\widehat{\Sl (2,\RR)}_{ce}$ is explained
by the existence of a {\em central extension} of the loop algebra,
with the {\em central charge} generator
\be
c := h_2 + h_3
\ee
This charge acts on the conformal factor $\lambda$ as a scaling
operator, but leaves all other fields inert \cite{Julia1,BM,Nic1}.
Finally, the trilinear Serre relations
\be\label{Serre2}
[f_2, [f_2 , [f_2 , f_3]]] = [f_3 ,[f_3 , [f_3 , f_2]]] = 0
\ee
are satisfied on all fields (the corresponding relations for the $e$ 
generators are trivially fulfilled). Together, \Ref{Serre1} and \Ref{Serre2} 
are just the {\em defining relations} (Chevalley Serre presentation 
\cite{Kac,MP,GO}) for the affine Lie algebra $A_1^{(1)}$.


Evidently, the relation \Ref{dualB0} between $\Bt$ and the Ehlers 
potential $B$ is a {\em nonlinear} extension of the duality
\be
\pa_\mu \varphi = \epsilon_{\mu\nu} \pa^\nu \tilde{\varphi}
\ee
valid for free scalar fields in two dimensions. The main difference is
that, whereas in the free field case, and more generally for $p$-form
gauge theories in higher dimensions, a second dualization brings us 
back to the field from which we started (modulo integration constants), 
iterating the duality transformations \Ref{Ehlers} and \Ref{MatznerMisner}
does not, as we pointed out already. {\em It is therefore the intrinsic 
non-linearity of Einstein's theory that explains the emergence of
an infinite chain of dualizations, and consequently of an infinite
dimensional symmetry.}

\section{Gravitational Billiards and Kac-Moody  Algebras}

The duality transformations reviewed in the previous section are 
invariances of mutilated versions of Einstein's theory only. On the 
other hand, what we are really after, are symmetries that would not 
require any such truncations. The symmetries we are about to discuss next 
considerably extend the ones discussed so far, but have not actually 
been shown to be symmetries of Einstein's theory, or some extension 
thereof. There are two reasons for this. First, the full gravitational 
field equations are far more complicated than the truncations discussed 
in the foregoing section --- as evidenced by the circumstance that no 
exact solutions appear to be known that would not make use of some 
kind of symmetry reduction in one way or another (in the appropriate
coordinates). Consequently, any extension of the known symmetries to the
full theory, which by necessity would be very non-local, will not be easy 
to identify. The second difficulty is that the Lie algebras that are 
conjectured to arise in this symmetry extension belong to the class
of {\em indefinite Kac Moody algebras}. However, after more than 
35 years of research in the theory of Kac Moody algebras, we still do not
know much more about these algebras beyond their mere existence --- despite 
the fact that they can be characterized by means of a simple set 
of generators and relations! \footnote{See remarks after table 1.1
to appreciate the challenge.} The main encouragement therefore derives
from the fact that there exists this link between these two seemingly 
unrelated areas, which provides more than just a hint of an as yet 
undiscovered symmetry of Einstein's theory. A key role in deriving 
these results was played by an analysis of Einstein's equations near 
a spacelike singularity in terms of {\em gravitational billiards},
to which we turn next.

\subsection{BKL dynamics and gravitational billiards}

A remarkable and most important development in theoretical cosmology 
was the analysis of spacelike (cosmological) singularities in Einstein's 
theory by Belinskii, Khalatnikov and Lifshitz (abbreviated as `BKL'
in the remainder), and their discovery of chaotic oscillations of
the spacetime metric near the initial singular hypersurface;\cite{BKL} 
see also \cite{BKL72,BKL2,Misner0,Misner1,Chitre}. There is a
large body of work on BKL cosmology, see \cite{Ryan,RyanSh,Jantzen} 
for recent reviews and extensions of the original BKL results. In particular,
there is now convincing evidence for the correctness of the basic BKL picture 
both from numerical analyses (see e.g. \cite{Berger,BerMon}) as 
well as from more rigorous work \cite{Cor,BeGaIs,AR,An}. It has also been 
known for a long time that the chaotic oscillations of the metric near the 
singularity can be understood in terms of {\em gravitational billiards}, 
although there exist several different realizations of this description,  
cf. \cite{Misner1,RyanSh,Jantzen} and references therein. The one which 
we will adopt here, grew out of an attempt to extend the original BKL results 
to more general matter coupled systems, in particular those arising in 
superstring and M theory\cite{DH1,DH2,DH3,DHJN,DaHeNi03}. It is particularly 
well suited for describing the relation between the BKL analysis and 
the theory of indefinite Kac Moody algebras, which is our main focus here, 
and which we will explain in the following section.
See also \cite{IvMe,Ivashchuk:1999rm} for an alternative approach.

We first summarize the basic picture, see \cite{DaHeNi03} for a 
more detailed exposition. Our discussion will be mostly heuristic, and 
we shall make no attempt at rigorous proofs here (in fact, the BKL 
hypothesis has been rigorously proven only with very restrictive assumptions
\cite{BeGaIs,BerMon,AR,An,DHRW}, but there is so far no proof of it in 
the general case). Quite generally, one considers a big-bang-like space-time 
with an initial singular spacelike hypersurface `located' at time $t=0$. 
It is then convenient to adopt a pseudo-Gaussian gauge for the metric 
(we will leave the number of spatial dimensions $d$ arbitrary for the moment)
\be
ds^2 = - N^2 dt^2 + g_{ij} dx^i dx^j
\ee
and to parametrize the spatial metric $g_{ij}$ in terms of a frame
field, or dreibein, $\theta^a$ (a one form) \footnote{The summation 
convention is in force for the coordinate indices $i,j,\dots$, but
suspended for frame indices $a,b,\dots$.}
\be
g_{ij} dx^i \otimes dx^j = \sum_{a=1}^d \theta^a \otimes \theta^a
\ee
For this frame field we adopt the so-called Iwasawa decomposition 
\be\label{Iwa}
\theta^a = e^{-\b^a}  {\cn^a}_i dx^i
\ee
by splitting off the (logarithmic) scale factors $\b^a$ from the
off-diagonal frame (and metric) degrees of freedom  ${\cn^a}_i$, which 
are represented by an upper triangular matrix with 1's on the diagonal. 
The spatial metric then becomes
\be
\label{Iwasawaex}
g_{ij} = \sum_{a=1}^d e^{- 2 \b^a}  {\cn^a}_i  \, {\cn^a}_j
\ee
The main advantage of the Iwasawa decomposition is that it matches 
precisely with the triangular decomposition (\ref{triangular}) below, which
is valid for any Kac Moody algebra. Furthermore, it turns out that, in 
the limit $t\rightarrow 0$ all the interesting action takes place 
in the scale factors $\b^a$, whereas the $\cn$ as well as the matter
degrees of freedom asymptotically `come to rest' in this limit.
Similarly, the metric and other degrees of freedom at {\em different
spatial points} should decouple in this limit, as the spatial distance
between them exceeds their horizon\footnote{One might even view this 
decoupling as a direct consequence of the {\em spacelike} nature of the 
singularity.}. The basic hypothesis underlying the BKL analysis is 
therefore that spatial gradients should become less and less important 
in comparison with time derivatives as $t\rightarrow 0$, such that
the resulting theory should be effectively describable in terms of a 
one dimensional reduction, in which the complicated partial differential 
equations of Einstein's theory are effectively replaced by a continuous 
infinity of ordinary differential equations.

To spell out this idea in more detail, let us insert the above metric 
ansatz into the Einstein-Hilbert action, and drop all spatial derivatives
(gradients), so that this action is approximated by {\em a continuous
superposition of one-dimensional systems}. One then obtains (still in
$d$ spatial dimensions)
\be\label{KasnerAction}
S[g_{ij}] = \frac14 \int d^d x \int dx^0 {\tilde{N}}^{-1} \left[ \left(\tr 
(\rg^{-1} \dot{\rg})^2 - (\tr \rg^{-1} \dot{\rg})^2\right)\right]
\ee
in a matrix notation where $\rg(t)\in GL(d,\RR)$ stands for 
the matrix $(g_{ij})$ representing the spatial components of the metric 
at each spatial point, and $\tilde{N}\equiv N\sqrt{g}$ is a rescaled lapse 
function. Neglecting the off-diagonal degrees of freedom, this action 
is further simplified to
\bqn\label{KasnerAction2}
S[\b^a] &=& \frac14 \int d^d x \int dx^0 {\tilde{N}}^{-1} 
\left[ \sum_{a=1}^{d} \big(\dot \b^a\big)^2 - 
     \left(\sum_{a=1}^{d} \dot \b^a \right)^2 \right]\non
&\equiv& \frac14 \int d^d x \int dx^0 {\tilde{N}}^{-1} 
G_{ab} \dot \b^a \dot \b^b
\eqn
where $G_{ab}$ is the restriction of the superspace metric ({\it \`a la} 
Wheeler-DeWitt) to the space of scale factors. A remarkable, and well known
property of this metric is its {\em indefinite signature} $(-+\cdots +)$,
with the negative sign corresponding to variations of the conformal
factor. This indefiniteness will be crucial here, because it directly 
relates to the indefiniteness of the generalized Cartan-Killing metric 
on the associated Kac Moody algebra. In the Hamiltonian description the 
velocities $\dot\b^a$ are replaced by their associated momenta $\pi_a$; 
variation of the lapse $\tilde{N}$ yields the {\em Hamiltonian constraint}
\be\label{Ham}
\ch = \sum_a \pi_a^2 - \frac1{d-1} \left(\sum_a \pi_a \right)^2
\equiv G^{ab} \pi_a \pi_b \approx 0 
\ee
Here $G^{ab}$ is the inverse of the superspace metric, i.e. $G^{ac} G_{bc}
= \delta^a_c$. The constraint \Ref{Ham} is supposed to hold at each spatial
point, but let us concentrate at one particular spatial point for the
moment. It is easy to check that \Ref{Ham} is solved by the well known 
conditions on the Kasner exponents. In this approximation, one thus has 
a Kasner-like metric at each spatial point, with the Kasner exponents 
depending on the spatial coordinate. 
In terms of the $\b$-space description, we thus have the following
picture of the dynamics of the scale factors at each spatial point.
The solution to the constraint \Ref{Ham} corresponds to the motion of 
a relativistic massless particle (often referred to as the `billiard 
ball' in the remainder) moving in the forward lightcone in $\b$-space 
along a lightlike line w.r.t. the `superspace metric' $G_{ab}$. 
The Hamiltonian constraint \Ref{Ham} is then re-interpreted as a 
relativistic dispersion relation for the `billiard ball'.

Of course, the above approximation does not solve the Einstein equations,
unless the Kasner exponents are taken to be constant (yielding the 
well known Kasner solution). Therefore, in a second step one must now 
take into account the spatial dependence and the effects of non-vanishing 
spatial curvature, and, eventually, the effect of matter couplings. 
At first sight this would seem to bring back the full complications 
of Einstein's equations. Surprisingly, this is not the case. Namely, 
one can show (at least heuristically) that\cite{DaHeNi03}

\begin{enumerate}
\item except for a finite number of them, the infinite number 
of degrees of freedom encoded in the spatially inhomogeneous metric, 
and in other fields, \textit{freeze} in that they tend to finite 
limits as $t\rightarrow 0$; and
\item the dynamics of the remaining `active' diagonal metric degrees of 
freedom can be asymptotically described in terms of a simple billiard 
dynamics taking place in the $\b$-space of (logarithmic) scale factors.
\end{enumerate}

This result can be expressed more mathematically as follows. In the 
limit $t\rightarrow 0$, the effect of the remaining degrees of freedom
consists simply in modifying the gravitational Hamiltonian \Ref{Ham} 
at a given spatial point by the addition of an {\em effective potential} 
that may be pictured as arising from `integrating out' all but the diagonal 
degrees of freedom. Accordingly, the free Hamiltonian constraint \Ref{Ham} 
is now replaced by an {\em effective Hamiltonian constraint}
\bqn
 \label{V4}
\ch(\b^a, \pi_{a},Q,P) =  G^{ab} \pi_a \pi_b  +
\sum_A c_A (Q,P)  \exp\big(- 2 w_A (\b)  \big) 
\eqn
where $\b^a,\pi_a$ are the canonical variables corresponding to the 
diagonal metric degrees of freedom, and $Q,P$ denote the remaining
canonical degrees of freedom. The quantities $w_A$ appearing in the 
exponential are generically {\em linear forms} in $\b$, 
\be
w_A (\b) = \sum_a (w_A)_a \b^a
\ee
and are usually referred to as {\em `wall forms'}. It is
crucial that the precise form of the coefficient functions $c_A(Q,P)$ 
--- which is very complicated --- does not matter in the BKL limit, 
which is furthermore dominated by a finite number of leading contributions 
for which $c_{A'}(Q,P)\geq 0$. The detailed analysis\cite{DaHeNi03} 
reveals various different kinds of walls: gravitational walls due to the 
effect of spatial curvature, symmetry (or centrifugal) walls resulting from 
the elimination of off-diagonal metric components, electric and magnetic 
$p$-form walls, and dilaton walls. It is another non-trivial result that 
all these walls are {\em timelike} in $\b$-space, that is, they have 
spacelike normal vectors.

The emergence of dominant walls is a consequence of the fact that, 
in the limit $t\rightarrow 0$, when $\b\rightarrow \infty$, most of 
the walls `disappear from sight', as the `soft' exponential walls become 
steeper and steeper, eventually rising to infinity. Perhaps a useful 
analogy here is to think of a mountainscape, defined by the sum of the 
exponential potentials $C_A e^{-2w_A}$; when the mountaintops rise into the 
sky, only the nearest mountains remain visible to the observer in the valley. 
The Hamiltonian constraint \Ref{V4} then takes the limiting form
\beq\label{V5}
H_{\infty}(\b^a, \pi_{a}) = G^{ab} \pi_a \pi_b  +
\sum_{A'}    \Theta_\infty \big(- 2 w_{A'} (\b)  \big)
\eeq
where the sum is only over the dominant walls (indexed by $A'$), and 
$\Theta_\infty$ denotes the infinite step function
\be
\Theta_\infty (x) := \left\{ \begin{array}{ll}
                      0  & \mbox{if $x<0$} \\
                      +\infty & \mbox{if $x>0$}
                      \end{array}
                      \right.
\ee 
In conclusion, the original Hamiltonian simplifies dramatically in the 
BKL limit $t\rightarrow 0$. The dynamics of \Ref{V5} is still that of a 
massless relativistic particle in $\b$-space, but one that is confined in a 
`box'. Hence, this particle undergoes occasional collisions with the `sharp' 
walls: when the argument of the $\Theta_\infty$ function is negative, i.e. 
between the walls, this particle follows a free relativistic motion 
characterized by the appropriate Kasner coefficients; when the particle 
hits a walls (where $\Theta_\infty$ jumps by an infinite amount), 
it gets reflected with a corresponding change in the Kasner exponents 
(these reflections are also referred to as {\em Kasner bounces}). Because 
the walls are timelike, the Kasner exponents get rotated by an element 
of the orthochronous Lorentz group in $\b$-space at each collision. 

In summary, we are indeed dealing with a {\em relativistic billiard} 
evolving in the forward lightcone in $\b$-space. The billiard walls 
(`cushions') are the hyperplanes in $\b$-space determined by the zeros 
of the wall forms, i.e. $w_{A'}(\beta)=0$. The chamber, in which the motion 
takes place, is therefore the wedge-like region defined by the inequalities  
\footnote{Note that this is a {\em space-time picture} in $\b$-space:
the walls {\em recede} as $t$ tends to 0, and $\b\rightarrow\infty$. 
The actual `billiard table' can be defined as the projection of this 
wedge onto the unit hyperboloid $G_{ab} \b^a \b^b = -1$ 
in $\beta$-space.\cite{DaHeNi03} See also \cite{Ryan,RyanSh,Jantzen}
for previous work and alternative descriptions of the billiard.}. 
\be\label{wA}
w_{A'} (\b) \geq 0
\ee

As for the long term (large $\b$) behavior of the billiard, there are 
two possibilities:\cite{DHJN} 

\begin{enumerate}
\item The chamber characterized by \Ref{wA} is entirely contained in
the forward lightcone in $\b$-space (usually with at least one edge
on the lightcone). In this case, the billiard ball will undergo infinitely 
many collisions because, moving at the speed of light, it will always 
catch up with one of the walls. The corresponding metric will then
exhibit infinitely many Kasner bounces between $0<t<\epsilon$ for
any $\epsilon >0$, hence {\em chaotic oscillations}. \footnote{Although 
we utilise this term in a somewhat cavalier manner here, readers can 
be assured that this system is indeed chaotic in the rigorous sense.
For instance, projection onto the unit hyperboloid in $\b$-space 
leads to a finite volume billiard on a hyperbolic manifold
of constant negative curvature, which is known to be strongly chaotic.}
\item The chamber extends beyond the lightcone, because some walls
intersect outside the lightcone. In this case the billiard ball undergoes
finitely many oscillations until its motion is directed towards a 
region that lies outside the lightcone; it then never catches up
with any wall anymore because no `cushion' impedes its motion.
The corresponding metrics therefore exhibit a {\em monotonic
Kasner-like behavior} for $0<t<\epsilon$ for sufficiently small
$\epsilon >0$.
\end{enumerate}

The question of chaotic vs. regular behavior of the metric near the
singularity is thereby reduced to determining whether the billiard
chamber realizes case 1 or case 2, and this is now a matter of a simple
algebraic computation. In the case of monotonic Kasner-like behavior
one can exploit these results and prove rigorous theorems about the 
behavior of the solution near the singularity.\cite{AR,DHRW} 

\subsection{Emergence of Kac Moody symmetries}

The billiard description holds not only for gravity itself, but
generalizes to various kinds of matter couplings extending the 
Einstein-Hilbert action. However, these billiards have no special
regularity properties in general. In particular, the dihedral angles 
between the `walls' bounding the billiard might depend on continuous 
couplings, and need not be integer submultiples of $\pi$. In some 
instances, however, the billiard can be identified with the fundamental 
Weyl chamber of a symmetrizable Kac Moody algebra of indefinite 
type\footnote{From now on we abbreviate `Kac Moody' by `KM',
and `Cartan subalgebra' by `CSA'.}, with Lorentzian signature 
metric.\cite{DH3,DHJN,DdBHS} Such billiards are also called `Kac Moody 
billiards'. Examples are pure gravity in any number of spacetime dimensions, 
for which the relevant KM algebra is $AE_d$, and superstring models\cite{DH3}
for which one obtains the rank $10$ algebras $E_{10}$ and $BE_{10}$, 
in line with earlier conjectures made in.\cite{Julia2} Furthermore, it
was understood that chaos (finite volume of the billiard) is equivalent 
to hyperbolicity of the underlying Kac Moody algebra.\cite{DHJN} 
Further examples of the emergence of Lorentzian Kac Moody algebras 
can be found in.\cite{DdBHS}.

The main feature of the gravitational billiards that can be associated
with KM algebras is that {\em there exists a group theoretical interpretation
of the billiard motion:} the asymptotic BKL dynamics is equivalent, at each 
spatial point, to the asymptotic dynamics of a one-dimensional nonlinear 
$\sigma$-model based on a certain infinite dimensional coset space $G/K(G)$, 
where the KM group $G$ and its maximal compact subgroup $K(G)$ depend on the 
specific model. In particular, {\em the $\beta$-space of logarithmic 
scale factors, in which the billard motion takes place, can be identified 
with the Cartan subalgebra (CSA) of the underlying indefinite Kac-Moody 
algebra.} The dominant walls that determine the billiards asymptotically
are associated with the simple roots of the KM algebra. We emphasize
that it is precisely the presence of gravity, which comes with an indefinite 
(Lorentzian) metric in the $\b$-superspace, hence a Cartan-Killing metric 
of indefinite signature, which forces us to consider {\it infinite 
dimensional} KM groups. By contrast, the finite dimensional simple Lie 
algebras, which can also be considered as KM algebras, but which were 
already classified long ago by Cartan, are characterized by a positive 
definite Cartan-Killing metric. 

The $\sigma$-model formulation to be introduced and elaborated in 
section~6 enables one to go beyond the BKL limit, and to see 
the beginnings of a possible identification of the dynamics of the 
scale factors {\em and} of all the remaining variables with that of a 
non-linear $\sigma$-model defined on the  cosets of the KM group divided by 
its maximal compact subgroup.\cite{DaHeNi02,DN} In that formulation, the 
various types of walls can thus be understood directly as arising from the 
large field limit of the corresponding $\s$-models. So far, only two examples 
have been considered in this context, namely pure gravity, in which case 
the relevant KM algebra is $AE_3$,\cite{DaHeNi03} and the bosonic 
sector of $D=11$ supergravity, for which the relevant algebra is the 
maximal rank 10 hyperbolic KM algebra $E_{10}$; we will return 
to the latter model in the final section. Following 
Ref.~\refcite{DaHeNi02,DN} one can introduce for both models a precise 
identification between the purely $t$-dependent $\s$-model quantities 
obtained from the geodesic action on the $G/K(G)$ coset space on 
the one hand, and the fields and their spatial gradients evaluated 
at a given, but arbitrarily chosen spatial point on the other.

\subsection{The main conjecture}

To sum up, it has been established that

\begin{enumerate}
\item in many physical theories of interest (and all the models arising 
in supergravity and superstring theory), the billiard region in which 
the dynamics of the active degrees of freedom takes place can be identified 
with the Weyl chamber of some Lorentzian KM algebra; and
\item the concept of a nonlinear $\s$-model on a coset space $G/K(G)$ can 
be generalized to the case where $G$ is a Lorentzian KM group, and
$K(G)$ its `maximal compact subgroup'; furthermore, these (one-dimensional) 
$\s$-models are asymptotically equivalent to the billiard dynamics 
describing the active degrees of freedom as $t\rightarrow 0$.
\end{enumerate}

So far, these correspondences between gravity or supergravity models
on the one hand, and certain KM coset space $\sigma$-models
on the other, work only for truncated versions of both models. Namely,
on the gravity side one has to restrict the dependence on the spatial
coordinates, whereas the KM models must be analyzed in terms
of a `level expansion', in which only the lowest orders are retained,
and the remaining vast expanse of the KM Lie algebra remains to be 
understood and explored. There are, however, indications that, at least 
as far as the higher order spatial gradients on the (super)gravity side 
are concerned, the correspondence can be further extended: the level 
expansions of $AE_3$, and other hyperbolic KM algebras 
contain all the requisite representations needed for the higher 
order spatial gradients\cite{DaHeNi02} (as well as an exponentially
increasing number of representations for which a physical interpretation 
remains to be found \cite{FisNic03}). This observation gave rise to 
the key conjecture\cite{DaHeNi02} for the correspondence between 
$D=11$ supergravity and the $\E/\KE$ coset model, which we here reformulate 
in a somewhat more general manner:

\begin{center}
\begin{flushleft}
{\it The time evolution of the geometric data at each spatial point, 
{\it i.e.} the values of all the fields and their spatial gradients,
can be mapped onto some constrained null geodesic motion on the infinite 
dimensional $G/K(G)$ coset space. }
\end{flushleft}
\end{center}

If true, this conjecture would provide us with an entirely 
new way of describing and analyzing a set of (non-linear) partial 
differential equations in terms of an ordinary differential equation in 
infinitely many variables, by `spreading' the spatial dependence over an 
infinite dimensional Lie algebra, and thereby mapping the cosmological
evolution onto a single trajectory in the corresponding coset space. 
In the remainder of this article we will therefore spell out
some of the technical details that lead up to this conjecture.

\section{Basics of Kac Moody theory}

We here summarize some basic results from the theory of KM
algebras, referring the reader to\cite{Kac,MP,GO} for comprehensive
treatments. Every KM algebra $\frakg \equiv \frakg (A)$ can 
be defined by means of an integer-valued Cartan matrix
$A$ and a set of generators and relations. We shall assume that 
the Cartan matrix is symmetrizable since this is the case encountered 
for cosmological billiards.  The Cartan matrix can then be written 
as ($i, j =1,\dots r $, with $r$ denoting the
{\it rank}
 of   $\frakg (A)$)
 \be A_{ij} = \frac{2   \langle
\a_i | \a_j \rangle }{\langle \a_i | \a_i \rangle }
\ee
where $\{\a_i  \}$ is a set of $r$
simple roots, and  where the angular brackets denote the invariant
symmetric bilinear form of $\frakg (A)$.\cite{Kac} Recall that the
roots can be abstractly defined as linear forms on the Cartan subalgebra
(CSA) $\frakh\subset\frakg (A)$. The generators, which are also 
referred to as Chevalley-Serre generators, consist of triples 
$\{h_i, e_i, f_i\}$ with $i=1,\dots,r$, and for each $i$ form 
an $\Sl(2,\RR)$ subalgebra. The CSA $\frakh$ is then spanned by the elements
$h_i$, so that 
\be 
[h_i, h_j] = 0 
\ee 
The remaining relations generalize the ones we already encountered
in Eqs. \Ref{Serre1} and \Ref{Serre2}:
Furthermore, 
\be 
[e_i , f_j]
= \d_{ij} h_j \ee and \be [h_i, e_j] =  A_{ij} e_j \; , \quad
[h_i, f_j] = - A_{ij} f_j 
\ee
so that the value of the linear form $\a_j$, corresponding to 
the raising operator $e_j$, on the element $h_i$  of the 
preferred basis $\{h_i \}$ of  $\frakh$ is $\a_j(h_i) = A_{ij}$.  
More abstractly, and independently of the choice
of any basis in the CSA,  the roots appear as eigenvalues of the
adjoint action of any element $h$ of the CSA on the raising ($e_i$) or
lowering  ($f_i$) generators: $[h,e_i] = + \a_i(h) e_i$,
$[h,f_i] = - \a_i(h) f_i$. Last but not least we have the 
so-called Serre relations
\be 
{\rm ad \,}(e_i)^{1-A_{ij}} \big(e_j\big) = 0 \; ,
\quad {\rm ad \,}(f_i)^{1-A_{ij}} \big(f_j\big) = 0 
\ee 

A key property of every KM algebra is the triangular decomposition 
\be\label{triangular} 
\frakg (A) = \frakn^- \oplus \frakh \oplus \frakn^+ 
\ee 
where $\frakn^+$ and
$\frakn^-$, respectively, are spanned by the multiple commutators
of the $e_i$ and $f_i$ which do not vanish on account of the Serre
relations or the Jacobi identity. To be completely precise,
$\frakn^+$ is the quotient of the free Lie algebra generated by
the $e_i$'s by the ideal generated by the Serre relations ({\it
idem} for $\frakn^-$ and $f_i$). In more mundane terms, when the
algebra is realized, in a suitable basis, by infinite dimensional
matrices, $\frakn^+$ and $\frakn^-$ simply consist of the `nilpotent' 
matrices with nonzero entries only above or below the diagonal. 
Exponentiating them formally, one obtains infinite dimensional matrices 
again with nonzero entries above or below the diagonal.

A main result of the general theory is that, for positive definite
$A$, one just recovers from these relations Cartan's list of finite
dimensional Lie algebras (see e.g.\cite{Humphreys} for an introduction). 
For non positive-definite $A$, on the other hand, the associated 
KM algebras are infinite dimensional. If $A$ has only one zero 
eigenvalue, with all other eigenvalues strictly positive, the associated 
algebra is called {\em affine}. The simplest example of such an algebra is 
the the $A_1^{(1)}$ algebra underlying the Geroch group, which we already 
encountered and discussed in section~2.2, with Cartan matrix \Ref{Geroch}. 
While the structure and properties of affine algebras
are reasonably well understood,\cite{Kac,GO} this is not so for 
{\em indefinite} $A$, when at least one eigenvalue of $A$ is negative. 
In this case, very little is known, and it remains an outstanding problem
to find a manageable representation for them.\cite{Kac,MP} In particular,
there is not a single example of an indefinite KM algebra for which
the root multiplicities, {\it i.e.} the number of Lie algebra elements
associated with a given root, are known in closed form. The scarcity of
results is even more acute for the `Kac-Moody groups' obtained by
formal exponentiation of the associated Lie algebras.
As a special, and important case, the class of  Lorentzian KM algebras 
includes {\it hyperbolic} KM algebras whose Cartan matrices are such 
that the deletion of any node from the Dynkin diagram leaves either 
a finite or an affine subalgebra, or a disjoint union of them.

The `maximal compact' subalgebra $\frakk$ is defined as the
invariant subalgebra of $\frakg (A)$ under the standard Chevalley
involution, {\it i.e.} \be \theta (x) = x  \quad {\rm for} \; \,
{\rm all} \;\, x\in\frakk \ee with \be \theta (h_i) = - h_i \; ,
\quad \theta (e_i) = - f_i \; , \quad \theta (f_i) = - e_i \ee
More explicitly, it is the subalgebra generated by multiple
commutators of $(e_i - f_i)$. For finite dimensional $\frakg (A)$,
the inner product induced on the maximal compact subalgebra
$\frakk$ is negative-definite, and the orthogonal complement to
$\frakk$ has a positive definite inner product. This is not true,
however, for indefinite $A$. It is sometimes convenient to introduce 
the operation of {\it transposition} acting on any Lie 
algebra element $E$ as 
\be
E^T := - \theta(E)
\ee
The subalgebra $\frakk$ is thus generated by the `anti-symmetric' 
elements satisfying $E^T = - E$; after exponentiation, the elements 
of the maximally compact subgroup $K$ formally appear as
`orthogonal matrices' obeying $k^T = k^{-1}$.

Often one uses a so-called Cartan-Weyl basis
for $\frakg(A)$. Using Greek indices $\mu,\nu,\dots$ to label the
root components corresponding to an arbitrary basis $H_{\m}$ in
the CSA, with the usual summation convention and a Lorentzian
metric $G_{\mu\nu}$ for an indefinite $\frakg$, we have $h_i :=
\a_i^\mu H_\mu$, where $ \a_i^\mu$ are the `contravariant
components', $G_{\mu\nu} \a_i^\n \equiv \a_{i \, \m}$, of the simple 
roots $\a_i$ ($i=1,\dots r$), which are linear forms on the CSA, with 
`covariant components' defined as $ \a_{i \, \m}  \equiv \a_i(H_\m)$.
To an arbitrary root $\a$ there corresponds a set of Lie-algebra generators
$E_{\a ,s}$, where $s=1, \dots, \mult (\a)$ labels the (in general) 
multiple Lie-algebra elements associated with $\a$. The root multiplicity
$\mult (\a)$ is always one for finite dimensional Lie algebras, and
also for the real (= positive norm) roots, but generically grows 
exponentiallly with $ -\a^2$ for indefinite $A$. 
In this notation, the remaining Chevalley-Serre generators are given 
by $e_i := E_{\a_i}$ and $f_i := E_{-\a_i}$. Then,
\be\label{HwE}
[ H_\mu , E_{\a,s} ] = \a_\mu E_{\a,s}
\ee
and
\be
[E_{\a,s} , E_{\a',t}] = \sum_u c_{\a\a'}^{s,t,u} E_{\a + \a',u}
\ee
The elements of the Cartan-Weyl basis are normalized such that
\be\label{CWnormalization}
\langle H_\mu | H_\nu \rangle = G_{\mu\nu} \; , \quad
\langle E_{\a,s} | E_{\b,t} \rangle = \d_{st} \d_{\a + \b, 0}
\ee
where we have assumed that the basis satisfies $E^T_{\a,s}  = E_{-\a,s}$.
Let us finally recall that the Weyl group of a KM algebra is the discrete group
generated by reflections in the hyperplanes orthogonal to the simple roots.

\section{The hyperbolic Kac Moody algebra $AE_3$}

As we explained, the known symmetries of Einstein's theory for
special types of solutions include the Ehlers and Matzner Misner 
$SL(2,\RR)$ symmetries, which can be combined into the Geroch group
$\widehat{SL(2,\RR)}_{ce}$. Furthermore, in the reduction to one time 
dimension, Einstein's theory is invariant under a rigid $SL(3,\RR)$ 
symmetry acting on the spatial dreibein. Hence, any conjectured symmetry 
of Einstein's theory should therefore contain these symmetries as
subgroups. Remarkably, there is a hyperbolic KM group with precisely 
these properties, whose Lie algebra is furthermore the simplest hyperbolic
KM algebra containing an affine subalgebra.\cite{Nic} This is the algebra 
$AE_3$, with Cartan matrix 
\be\label{AE3} 
\hspace{1.5cm}
A_{ij} = \left( \begin{matrix} 
2 &-1 & 0 \cr
-1 & 2 & -2 \cr
0 & -2 & 2  
\end{matrix} \right)
\ee
The $\Sl(2,\RR)$ subalgebra corresponding to the third diagonal entry 
of $A_{ij}$ is associated with the Ehlers group. The affine subgroup 
corresponding to the submatrix \Ref{Geroch} is the Geroch 
group\cite{Geroch}  already discussed in section~2.2. The $SL(3,\RR)$ 
subgroup containing the the Matzner-Misner $SL(2,\RR)$ group, is generated 
by $(e_1,f_1,h_1)$ and $(e_2,f_2,h_2)$, corresponding to the submatrix
\be\label{SL3}
\hspace{1.5cm} \left( 
\begin{matrix}2 &-1\cr -1 & 2  
\end{matrix} \right)
\ee

As we said, not much is known about $AE_3$; in particular, there is 
no `list' of its (infinitely many) generators, nor of its structure 
constants (which are certainly too numerous to fit in any list, see below!). 
Nevertheless, in order to gain some `feeling' for this algebra, we 
will now work out the beginnings of its decomposition into irreducible 
representations of its $SL(3,\RR)$ subgroup. Of course,
this decomposition refers to the {\it adjoint action} of the $\Sl(3,\RR)$  
subalgebra embedded in $AE_3$. More specifically, we will analyze the 
lowest terms of the nilpotent subalgebra $\frakn^+$. To do so, we 
first define, for any given root $\a$, its $\mathfrak{sl}(3,\RR)$ level 
$\ell$ to be the number of times the root $\a_3$ appears in it, to wit,
$\a = m \a_1 + n \a_2 + \ell \a_3$.
The algebra $AE_3$ thereby decomposes into an infinite
irreducible representations of its $\mathfrak{sl}(3,\RR)$ 
subalgebra\footnote{A different decomposition would be one in terms of 
the affine subalgebra $A_1^{(1)} \subset AE_3$; \cite{FF} however, 
the representation theory of $A_1^{(1)}$ is far more complicated and 
much less developed than that of $\mathfrak{sl}(3,\RR)$.}.
As is well known,\cite{Humphreys} the irreducible representations 
of $\Sl(3,\RR)$ can be conveniently characterized by their Dynkin labels 
$[p_1,p_2]$. In terms of the Young tableau description of $\Sl(3,\RR)$ 
representations, the first Dynkin label $p_1$ counts the number of 
columns having  two boxes, while $p_2$ counts the number of columns 
having only one box. For instance, $[p_1,p_2] = [1,0]$ labels an 
antisymmetric two-index tensor, while $[p_1,p_2] = [0,2]$ denotes a 
symmetric two-index tensor. The dimension of the representation 
$[p_1,p_2]$ is $ ( p_1 +1) ( p_2 +1) ( p_1 +p_2 +2) / 2 $.

The level $\ell =0$ sector, which includes the third Cartan generator 
$h_3$, is the $\mathfrak{gl}(3,\RR)$ subalgebra with generators 
${K^i}_j$ (where $i,j =1,2,3$) and commutation relations
\be
[{K^i}_j , {K^k}_l ] =  \d^k_j {K^i}_l -\d^i_l  {K^k}_j
\ee
corresponding to the $GL(3,\RR)$ group acting on the spatial components 
of the vierbein. The restriction of the $AE_3$-invariant bilinear form 
to the level-0 sector is
\be
\label{KK}
\langle {K^i}_j | {K^k}_l \rangle = \d^i_l \d^k_j - \d^i_j \d^k_l
\ee
The identification with the Chevalley-Serre generators is
\beq
e_1 &=& {K^1}_2 \;\; ,  \quad f_1 = {K^2}_1 \;\; ,  \quad
       h_1 = {K^1}_1 - {K^2}_2  \nn\\
e_2 &=& {K^2}_3 \;\; ,  \quad f_2 = {K^3}_2 \;\; ,  \quad
       h_2 = {K^2}_2 - {K^3}_3  \nn\\
h_3 &=& -  {K^1}_1 - {K^2}_2 + {K^3}_3
\eeq
showing how the over-extended CSA generator $h_3$ enlarges the
original  $\Sl(3,\RR)$  generated by $(e_1,f_1,h_1)$ and  $(e_2,f_2,h_2)$  
to the Lie algebra $\mathfrak{gl}(3,\RR)$. The CSA generators are related to
the `central charge' generator $c$ by
\be\label{centralcharge}
c= h_2 + h_3 = - {K^1}_1
\ee
which acts as a scaling on the conformal factor\cite{Julia1,BM,Nic1} 
(here realized as the 1-1 component of the vierbein). 

To determine the representations of $\Sl(3,\RR)$ appearing at
levels $\ell = \pm 1$, we observe that, under the adjoint action  
of $\Sl(3,\RR)$, i.e. of $(e_1,f_1,h_1)$ and $(e_2,f_2,h_2)$, 
the extra Chevalley-Serre generator $f_3$ is a highest weight vector:
\beq
e_1 (f_3) \equiv [e_1, f_3] &=& 0 \non
e_2 (f_3) \equiv [e_2, f_3] &=& 0
\eeq
The Dynkin labels of the representation built on this highest weight
vector $f_3$ are $(p_1,p_2) = (0,2)$, since
\beq
h_1 (f_3)\equiv [h_1, f_3] &=& 0 \non
h_2 (f_3)\equiv [h_2, f_3] &=& 2  f_3
\eeq
As mentioned above, the representation $(p_1,p_2) = (0,2)$ corresponds to
a symmetric (two-index) tensor. Hence, at the levels $\pm 1$ we have 
$AE_3$ generators which can be represented as  symmetric tensors 
$E^{ij}= E^{ji}$ and $F_{ij}= F_{ji}$. One verifies that all algebra 
relations are satisfied with
($a_{(ij)} \equiv (a_{ij} +a_{ji}) /2$)
\beq\label{AE3com}
[{K^i}_j, E^{kl}] &=& \d^k_j E^{il} + \d^l_j E^{ki} \non
{}[{K^i}_j, F_{kl}] &=&  - \d^i_k F_{jl} - \d^i_l F_{kj} \non
{}[E^{ij} , F_{kl} ] &=& 2 \d^{(i}_{(k} {K^{j)}}_{l)} -
  \d^{(i}_{k} \d^{j)}_{l} \big({K^1}_1 + {K^2}_2 + {K^3}_3 \big)\non
\langle F_{ij} | E^{kl}\rangle &=&  \d^{(k}_{i} \d^{l)}_{j}
\eeq
and the identifications
\be
e_3 = E^{33} \;\; , \quad f_3 = F_{33}
\ee

As one proceeds to higher levels, the classification of $\Sl(3,\RR)$
representations becomes rapidly more involved due to the exponential
increase in the number of representations with level $\ell$.
Generally, the representations that can occur at level $\ell + 1$
must be contained in the product of the level-$\ell$ representations
with the level-one representation $(0,2)$. Working out these products
is elementary, but cumbersome. For instance, the level-two generator
$ E^{ab | jk} \equiv \varepsilon^{abi} {E_i}^{jk}$,
with labels $(1,2)$,
is straightforwardly obtained by commuting two level-one elements
\be
[ E^{ij} , E^{kl} ] =  \varepsilon^{mk(i} {E_m}^{j)l} +
                       \varepsilon^{ml(i} {E_m}^{j)k}
\ee
A more economical way to identify the relevant representations
is to work out the relation between Dynkin labels and the associated
highest weights, using the fact that the highest weights of the adjoint 
representation are the roots.  More precisely, the highest weight
vectors being (as exemplified above at level 1) of the `lowering type',
the corresponding highest weights are {\it negative} roots, say 
$\Lambda =  -\a$. Working out the associated Dynkin labels one obtains
\be
p_1 \equiv p = n -2m \;\; , \quad p_2 \equiv q = 2\ell + m - 2n
\ee
As indicated, we shall henceforth use the notation $ [p_1,p_2] \equiv [p,q]$
for the Dynkin labels. This formula is restrictive because all the integers
entering it must be non-negative. Inverting this relation we get
\beq\label{mn}
m &=& \ft23 \ell - \ft23 p - \ft13 q  \non
n &=& \ft43 \ell - \ft13 p - \ft23 q
\eeq
with $n \geq 2m \geq 0$. A further restriction derives from
the fact that the highest weight must be a root of $AE_3$, viz.
its square must be smaller or equal to 2:
\be\label{Lambda2}
\Lambda^2 = \ft23 \big( p^2 + q^2 + pq - \ell^2 \big) \leq 2
\ee
Consequently, the representations occurring at level $\ell$ must belong
to the list of all the solutions of \Ref{mn} which are
such that the labels $m,n,p,q$ are non-negative integers and
the highest weight $\Lambda$ is a root, i.e. $\Lambda^2 \leq 2$.
These simple diophantine equations/inequalities can be easily
evaluated by hand up to rather high levels.

Although the above procedure substantially reduces the number of 
possibilities, it does not tell us how often a given representation 
appears, i.e. its {\it outer multiplicity} $\mu$. For this purpose we have 
to make use of more detailed information about $AE_3$, namely the root
multiplicities computed in.\cite{FF,Kac} Matching the combined
weight diagrams with the root multiplicities listed in table $H_3$
on page 215 of,\cite{Kac} one obtains the following representations
in the decomposition of $AE_3$ w.r.t. its $\Sl(3,\RR)$ subalgebra up
to level $\ell\leq 5$, where we also indicate the root coefficients
$(m_1,m_2,\ell)$, the norm and multiplicity of the root $\alpha$,
and the outer multiplicity of the representation $[p,q]$: 

\vspace*{0.4cm}
\begin{longtable}{|c|c|c|c|c|c|}
\hline
$\ell$&$[p,q]$&$\a$&$\a^2$&$\textrm{mult}\,\a$&$\m$\\
\hline
\hline
\endhead
1&[0,2]&(0,0,1)&2&1&1\\
2&[1,2]&(0,1,2)&2&1&1\\
3&[2,2]&(0,2,3)&2&1&1\\
&[1,1]&(1,3,3)&-4&3&1\\
4&[3,2]&(0,3,4)&2&1&1\\
&[2,1]&(1,4,4)&-6&5&2\\
&[1,0]&(2,5,4)&-10&11&1\\
&[0,2]&(2,4,4)&-8&7&1\\
&[1,3]&(1,3,4)&-2&2&1\\
5&[4,2]&(0,4,5)&2&1&1\\
&[3,1]&(1,5,5)&-8&7&3\\
&[2,0]&(2,6,5)&-14&22&3\\
&[0,1]&(3,6,5)&-16&30&2\\
&[0,4]&(2,4,5)&-6&5&2\\
&[1,2]&(2,5,5)&-12&15&4\\
&[2,3]&(1,4,5)&-4&3&2\\
\hline\\
\caption{\label{sl3cont}\sl Decomposition of $AE_3$ under $\Sl(3,\RR)$ 
for $\ell \leq 5$.}
\end{longtable}

The above table does not look too bad, but appearances are deceptive,
because the number of representations grows exponentially with the level!
For $AE_3$, the list of representations 
with their outer multiplicities is meanwhile available up to $\ell \leq 56$ 
\cite{FisNic03}; the total number of representations up to that level is 
$20\,994\,472\,770\,550\,672\,476\,591\,949\,725\,720$ 
\footnote{T.~Fischbacher, private communication.}, larger than $10^{31}$! 
This number should suffice to convince readers of the `explosion' that 
takes place in these algebras as one increases the level. Similar 
decompositions can be worked out for the indefinite Kac-Moody algebras 
$E_{10}$ and $E_{11}$ \cite{FisNic03}, and for $E_{10}$ under its 
$D_9$ and $A_8\times A_1$ subalgebras.\cite{KleNic04,KleNicB}. The real
problem, however, is not so much the large number of representations, 
but rather the absence of any discernible structure in these tables,
at least up until now.

\section{Nonlinear $\sigma$-Models in one dimension}

Notwithstanding the fact that we know even less about the groups
associated with indefinite KM algebras, it is possible to formulate 
nonlinear $\s$-models in one time dimension and thereby provide an
effective and unified description of the asymptotic BKL dynamics
for several physically important models. The basic object of interest 
is a one-parameter dependent KM group element $\cV =\cV(t)$, assumed to 
be an element of the coset space $G/K(G)$, where $G$ is the group 
obtained by formal exponentiation of the KM algebra $\frakg$, and 
$K(G)$ its maximal compact subgroup, obtained by formal exponentiation
of the associated maximal compact subalgebra $\frakk$ defined above.
For finite dimensional $\frakg(A)$ our definitions reduce to the usual
ones, whereas for indefinite KM algebras they are formal constructs
to begin with. In order to ensure that our definitions are meaningful
operationally, we must make sure at every step that any finite truncation
of the model is well defined and can be worked out explicitly in a
finite number of steps.

In physical terms, $\cv$ can be thought of as an extension of the
vielbein of general relativity, with $G$ and $K(G)$ as generalizations
of the $GL(d,\RR)$ and local Lorentz symmetries of general relativity.
For infinite dimensional $G$, the object $\cv$ thus is a kind of
`$\infty$-bein', that can be associated with the `metric'
\be
\cm := \cv^T \cv
\ee
which is invariant under the  left action ( $\cv \to  k \cv$)
of the `Lorentz group' $K(G)$. Exploiting this invariance, 
we can formally bring $\cv$ into a triangular gauge
\be\label{Iwasawa3}
\cv = \ca \cdot \cn \Longrightarrow \cm = \cn^T \ca^2 \cn
\ee
where the abelian part $\ca$ belongs to the exponentiation of the CSA,
and the nilpotent part $\cn$ to the exponentiation  of $\frakn^+$.
This formal Iwasawa decomposition, which is the infinite dimensional
analog of \Ref{Iwa}, can be made fully explicit 
by decomposing $\ca$ and  $\cn$ in terms of bases of  
$\frakh$ and $\frakn^+$ (using the Cartan Weyl basis)
\beq\label{Iwasawa4}
\ca (t) &=& \exp \big( \b^\mu (t) \, H_\mu \big) \; , \non
\cn (t) &=& \exp \Big( \SD \Smult \nu_{\a ,s} (t) \, E_{\a , s}\Big)
\eeq
where $\Delta_+$ denotes the set of positive roots. The components    
$ \b^\mu $, parametrizing a generic element in the CSA  $\frakh$,
will turn out to be in direct correspondence with the metric scale 
factors $ \b^a$ in \Ref{Iwasawaex}. The main technical difference 
with the kind of Iwasawa decompositions used in section~3.1 is that 
now the matrix $\cv (t)$ is infinite dimensional for indefinite 
$\frakg (A)$, in which case the decomposition \Ref{Iwasawa4} is, 
in fact, the only sensible parametrization available! Consequently, 
there are now infinitely many $\nu$'s, whence $\cn$ contains 
an infinite tower of new degrees of freedom. Next we define
\be\label{Ndot}
\dot \cn  \cn^{-1} =  \SD\Smult j_{\a ,s} E_{\a,s} \quad \in \frakn^+
\ee
with
\be
j_{\a ,s} = \dot \nu_{\a ,s} + ``\n \dot \n + \n \n \dot \n  + \cdots''
\ee
(we put quotation marks to avoid having to write out the indices).
To define a Lagrangian we consider the quantity
\be
\dot\cv\cv^{-1} = \dot\b^\mu H_\mu +
    \SD\Smult \exp\big(\a (\b)\big)  j_{\a ,s} E_{\a,s}
\ee which has values in the Lie algebra $\frakg (A)$. Here we have
set \be \a (\b) \equiv \a_\mu \b^\mu \ee for the value of the root
$\a$ ( $\equiv$ linear form) on the CSA element  $\b = \b^\m
H_\m$. 
Next we define 
\beq 
P&:=& \frac12 \left( \dot\cv\cv^{-1} +
(\dot\cv\cv^{-1})^T \right) \nonumber \\ 
&=&   \dot\b^\mu H_\mu +
\frac12 \SD\Smult j_{\a ,s} \exp\big(\a (\b)\big) (E_{\a,s} + E_{-\a,s})
\eeq
where we arranged the basis so that $E_{\a,s}^T =  E_{-\a,s}$. 
The KM-invariant $\s$-model Lagrangian is defined by means of the
KM-invariant bilinear form 
\beq\label{Lag}
\cl &=& \frac12 n^{-1} \langle P | P \rangle \non
&=& n^{-1} \Big(\frac12 G_{\m\n}  \dot\b^\mu \dot\b^\nu +
   \frac14 \SD\Smult \exp\big(2\a(\b)\big) 
   j_{\a,s} j_{\a,s}\Big) \label{Lagrangian}
\eeq
Here the Lorentzian metric $ G_{\m\n}$ is the restriction of the 
invariant bilinear form to the CSA, cf. \Ref{CWnormalization}. 
The `lapse function' $n$ ensures that our formalism is invariant 
under  reparametrizations of the time variable. Remarkably, this 
action defined by the above Lagrangian is  {\em essentially unique}
because there are no higher order polynomial invariants for indefinite
KM algebras.\cite{Kac}

After these preparations we are now ready to specialize to the 
algebra $AE_3$. In this case this Lagrangian \Ref{Lag} contains
the Kasner Lagrangian (\ref{KasnerAction}) as a special truncation. 
More specifically, retaining only the level zero fields 
(corresponding to the `sub-coset' $GL(3,\RR)/O(3)$)
\be
\cv(t) \Big|_{\ell =0} =   \exp (  {h^a}_b(t)  {K^b}_a )
\ee
and defining from $ {h^a}_b$ a vielbein by matrix exponentiation
$ {e^a}_b \equiv {(\exp h)^a}_b$, and a corresponding contravariant 
metric $g^{ab} = {e^a}_c {e^b}_c$, it turns out that the bilinear 
form (\ref{KK}) reproduces the Lagrangian (\ref{KasnerAction}) (for
the special case of three spatial dimensions). This means that
{\em we can identify the restriction $G_{\mu\nu}$ of the Cartan-Killing
metric to the CSA with the superspace metric $G_{ab}$ in the superspace 
of scale factors $\b$ in \Ref{KasnerAction}.} 

At level $\ell=1$, we have the fields $\phi_{ij}$ associated with the 
level-one generators $E^{ij}$. Observe that for $D=4$, these are precisely
the spatial components of the dual graviton introduced in \Ref{dualG} ---
in other words, we have rederived the result of section 2.2 by a 
purely group theoretical argument! (This argument works likewise for
$D>4$.) This leads to a slightly less restricted truncation of our 
KM-invariant $\s$-model
 \be
\cv(t) \Big|_{\ell =0,1} = \exp ({h^a}_b(t) {K^b}_a ) \exp(\phi_{ab} E^{ab})
\ee
In the gauge $n=1$, the Lagrangian now has the form
$\cl \sim (g^{-1} \dot g  )^2 + g^{-1} g^{-1} \dot \phi \dot \phi$,
where $g$ denotes the covariant metric $g_{ij}$. As the $\phi_{ij}$'s 
enter only through their time derivatives, their conjugate momenta 
$\Pi^{ij}$ are constants of the motion in this $|\ell|\leq 1$
truncation. Eliminating the $\phi$'s in terms of the constant momenta 
$\Pi$ yields 
\be
V_{\phi}(g) \propto + g_{ij} g_{kl} \Pi^{ik} \Pi^{jl}
\ee
This potential can be identified with the leading (weight-2) gravitational 
potential, if we identify
the structure constants ${C^i}_{jk}$ defined by $d\theta^i =
{C^i}_{jk} \theta^j\wedge \theta^k$, 
with the momenta conjugate to $\phi_{ij}$ as
\be
\Pi^{ij} = \varepsilon^{kl(i} {C^{j)}}_{kl}
\ee
Consequently, the BKL dynamics at each spatial point is equivalent 
to the $|\ell|\leq  1$ truncation of the $AE_3$-invariant dynamics 
defined by \Ref{Lagrangian}. The  fields $\phi_{ij}(t)$ parametrizing 
the components of the $AE_3$ coset element along the $\ell =1$ generators 
are canonically conjugate to the structure constants ${C^i}_{jk}$.
The proper physical interpretation of the higher level fields remains 
yet to be found. 

Varying (\ref{Lagrangian}) w.r.t. the lapse function $n$ gives rise to
the constraint that the coset Lagrangian vanish. Defining the 
canonical momenta
\be
\pi_a := \frac{\d\cl}{\d\dot \b^a} = n^{-1} G_{ab} \dot\b^b
\ee
and the (non-canonical) momentum-like variables
\be
\Pi_{\a,s} := \frac{\d\cl}{\d j_{\a,s}}
      = \frac12 n^{-1} \exp \big(2\a(\b)\big)j_{\a,s}
\ee
and recalling the equivalence of the Cartan Killing and superspace metrics
noted above, we are led to the Hamiltonian constraint of the $\sigma$-model,
which is given by
\beq\label{HKM}
\ch(\b^a,\pi_a,...) =  \frac12  G^{ab} \pi_a \pi_b
      + \sum_{\a\in\Delta_+} \sum_{s=1}^{{\rm mult}(\a)} 
\exp\big(-2\a(\b)\big) \Pi_{\a ,s} \Pi_{\a ,s}\qquad
\eeq
where $\b^a,\pi_a$ are now the diagonal CSA degrees of freedom,
and the dots stand for infinitely many off-diagonal (Iwasawa-type)
canonical variables, on which the $\Pi_{\a,s}$ depend. 

The evident similarity of \Ref{V4} and \Ref{HKM} is quite striking, 
but at this point we can only assert that the two expressions coincide 
asymptotically, when they both reduce to a relativistic billiard.
Namely, because the coefficients of the exponentials in \Ref{HKM} are
non-negative, we can apply exactly the same reasoning as for
the gravitational billiards in section~3.1. One then finds that 
the off-diagonal components $\nu_{\a,s}$ and the momentum-like variables 
$\Pi_{\a,s}$ get frozen asymptotically (again, we may invoke the
imagery of a mountainscape, now defined by exponential potentials 
for all roots). In the present KM setup, all the walls enter on the 
same footing; there is nothing left of the distinctions between 
different types of walls (symmetry, gravitational, electric, and so
on). The only important characteristic of a wall is its height 
${\rm ht} \, \a \equiv n_1 + n_2 +\cdots$ for a root decomposed 
along simple roots as $ \a = n_1 \a_1 + n_2 \a_2 + \cdots$.
The asymptotic Hamiltonian hence is dominated by the walls associated 
to the {\em simple roots}:
\be\label{HC2} 
\ch_\infty (\b, \pi) = \frac12 \pi^a \pi_a +
  \sum_{i=1}^r  \, \Theta_\infty \big(-2\a_i (\b)\big)
\ee
where the sum is over the simple roots only, and the
motion of the $\b^a$ is confined to the fundamental Weyl
chamber $\a_i(\b) \geq 0$. 

The billiard picture for pure gravity in four dimensions is now
readily understood in terms of the Weyl group of $AE_3$,\cite{DHJN} 
which is just the modular group $PGL(2,\ZZ)$,\cite{FF} and the 
simple roots of $AE_3$. For the $\Sl(3,\RR)$ subalgebra, which has 
two simple roots, the Weyl group is the permutation group on three 
objects. The two hyperplanes orthogonal to these simple roots can be
identified with the symmetry (centrifugal) walls. The third simple 
root extending (\ref{SL3}) to the full rank 3 algebra (\ref{AE3}) 
can be identified the dominant curvature (gravitational) wall.

To conclude: {\em in the limit where one goes to infinity in the Cartan 
directions, the dynamics of the Cartan degrees of freedom of the coset 
model become equivalent to a billiard motion within the Weyl chamber, 
subject to the zero-energy constraint  $\ch_\infty (\b, \pi) =0$.}
Therefore, in those cases where the gravitational billiards of section~3.1
are of KM-type, they are asymptotically equivalent to the KM $\sigma$-models 
over $G/K(G)$.

\section{Finale: $\E$ -- the ultimate symmetry?}

There can be little doubt that the algebra, which from many points is 
the most intriguing and most beautiful, is the maximal rank hyperbolic 
KM algebra $\E$, which is an infinite dimensional extension of the
better known finite dimensional exceptional Lie algebras $E_6, E_7$ 
and $E_8$. \cite{Humphreys} There are two other rank-10 
hyperbolic KM algebras $DE_{10}$ and $BE_{10}$ (respectively related to 
type I supergravity, and Einstein Maxwell supergravity in ten 
dimensions), but they appear to be less distinguished. The emergence
of $\E$ in the reduction of $D=11$ supergravity to one dimension had 
first been conjectured in.\cite{Julia2} A crucial new feature of the 
scheme proposed here, which is based on a hyperbolic $\sigma$-model 
defined by means of the geodesic action \Ref{Lag} is that it retains 
a residual spatial dependence, which on the $\s$-model side is supposed 
`to be spread' over the whole $\E$ Lie algebra. Thereby all degrees 
of freedom of the original theory should still be there, unlike for 
a {\it bona fide} reduction to one dimension. 

Just like $AE_3$ the KM algebra $\E$ algebra is recursively defined via 
its Chevalley-Serre presentation in terms of generators and relations 
and its Dynkin diagram which we give below. 

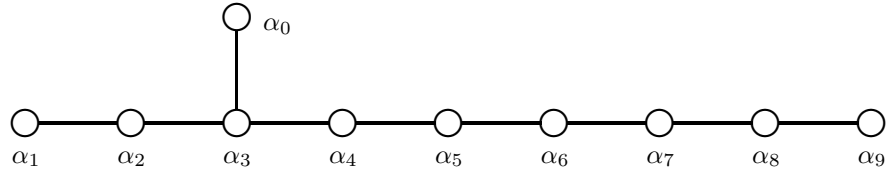
\begin{center}
\scalebox{1}{
\begin{picture}(340,60)
\put(5,-5){$\alpha_1$}
\put(45,-5){$\alpha_2$}
\put(85,-5){$\alpha_3$}
\put(125,-5){$\alpha_4$}
\put(165,-5){$\alpha_5$}
\put(205,-5){$\alpha_6$}
\put(245,-5){$\alpha_7$}
\put(285,-5){$\alpha_8$}
\put(325,-5){$\alpha_9$}
\put(100,45){$\alpha_{0}$}
\thicklines
\multiput(10,10)(40,0){9}{\circle{10}}
\multiput(15,10)(40,0){8}{\line(1,0){30}}
\put(90,50){\circle{10}} \put(90,15){\line(0,1){30}}
\end{picture}}\end{center}
\vspace*{0.4cm}
The nine simple roots $\a_1,\dots,\a_9$ along the horizontal line 
generate an $A_9 \equiv \Sl(10,\RR)$ subalgebra. One of the reasons 
why $\E$ is distinguished is that its root lattice is the unique even 
self-dual Lorentzian lattice II$_{1,9}$ (such lattices exist only in
dimensions $d=2 +8n$.\cite{CS})

For the corresponding $\sigma$-model a precise identification can be made 
between the purely $t$-dependent $\s$-model quantities obtained from the 
geodesic action on the $\E/\KE$ coset space on the one hand, and certain 
fields of $D=11$ supergravity and their spatial gradients evaluated at a 
given, but arbitrarily chosen spatial point on the other.\cite{DaHeNi02,DN} 
The simple and essentially unique geodesic Lagrangian describing a null 
world line in the infinite-dimensional coset manifold $E_{10}/K(E_{10})$ 
thus reproduces the dynamics of the bosonic sector of eleven-dimensional 
supergravity in the vicinity of a space-like singularity. This result 
can be extended to {\em massive} ${\rm IIA}$ supergravity,\cite{KleNic04} 
where also parts of the fermionic sector were treated for the first time, 
and to IIB supergravity in.\cite{KleNicB} Related results had been previously
obtained in the framework of $E_{11}$.\cite{SWest,SWest2,Kleinschmidt:2003mf} 

A main ingredient in the derivation of these results is the level 
decomposition of $E_{10}$ w.r.t. the $A_9$, $D_9$, and $A_8\times A_1$
subalgebras of $E_{10}$, respectively, which generalizes the
$\Sl(3,\RR)$ decomposition of $AE_3$ made in section~5. In all cases,
one obtains precisely the field representation content of the 
corresponding supergravity theories at the lowest levels, and for
all these decompositions, the bosonic supergravity equations of motion, 
when restricted to zeroth and first order spatial gradients, match with 
the corresponding $\s$-model equations of motion at the lowest levels.
In particular, the self-duality of the five-form field strength 
in type IIB supergravity is implied by the dynamical matching between 
the  $E_{10}/K(E_{10})$ $\sigma$-model and the supergravity equations 
of motion, and does not require local supersymmetry or some other 
extraneous argument for its explanation. 

Combining the known results, we can summarize the correspondence 
between the maximally supersymmetric theories and the maximal rank 
regular subalgebras of $E_{10}$ as follows
\begin{eqnarray}
A_9 \subset E_{10} \quad &\Longleftrightarrow& \qquad
  \mbox{$D=11$ supergravity} \nn\\
D_9 \subset E_{10} \quad &\Longleftrightarrow& \qquad
  \mbox{massive IIA supergravity} \nn\\
A_8 \times A_1 \subset E_{10} \quad &\Longleftrightarrow& \qquad
  \mbox{IIB supergravity} \nn
\end{eqnarray}
The decompositions of $E_{10}$ w.r.t. its other rank 9 regular subalgebras
$A_{D-2} \times E_{11-D}$ (for $D=3,\ldots,9$) will similarly reproduce 
the representation content of maximal supergravities in $D$ space-time 
dimensions as the lowest level representations. 

We conclude by repeating the main challenge that remains: one must 
extend these correspondences to higher levels and spatial gradients, 
and find a physical interpretation for the higher level representations, 
whose number exhibits an exponential growth similar to the growth in 
the number of excited string states (see, however, \cite{DN05} for recent
progress concerning the link between higher order M Theory corrections
and the $E_{10}$ root lattice). Because this will inevitably 
require (or entail) a detailed understanding of indefinite and
hyperbolic KM algebras, it might also help in solving the core problem 
of the theory of Kac Moody algebras, a problem that has vexed 
almost a generation of researchers.

\section*{Acknowledgments} 
This work was supported in part by the European Network
HPRN-CT-2000-00122 and by the the German Israeli Foundation (GIF)
Project Nr. I. 645 130-14--1999. It is a great pleasure to thank 
T. Damour, T. Fischbacher, M. Henneaux and A. Kleinschmidt
for enjoyable collaborations and innumerable discussions that have 
shaped my understanding of the results reported here, and I.H.E.S.,
Bures-sur-Yvette, for continued support during several visits there.
I am also very grateful to F. Englert, A. Feingold, A. Giveon, L. Houart 
and E. Rabinovici for enlightening discussions at various earlier 
stages of this work.

\end{document}